\newif\ifcomments\commentstrue
\documentclass[letterpaper,twocolumn,10pt]{article}
\usepackage{usenix2019_v3}

\usepackage{tikz}
\usepackage{amsmath}

\usetikzlibrary{positioning,matrix,fit,shapes,backgrounds,patterns,arrows, arrows.meta, calc}

\usepackage{pgfplots}
\pgfplotsset{
	every axis/.prefix style={
		label style={font=\footnotesize},
		tick label style={font=\footnotesize, 
			scaled y ticks = false,
			 legend style={font=\footnotesize},
		},
	},
	every axis plot/.append style={thick},
	compat = 1.3,
}

\usepackage{amsmath}
\PassOptionsToPackage{hyphens}{url}
\usepackage{cleveref}  

\usepackage{xspace}

\usepackage{multirow}

\usepackage{xcolor}
\usepackage{amssymb}
\usepackage{multirow}
\usepackage{subfig}
\usepackage{algorithm}
\usepackage[noend]{algpseudocode}
\usepackage{xspace}
\usepackage{svg}
\usepackage{cite}
\usepackage{bbding}
\usepackage{xurl}
\usepackage[frozencache,cachedir=.]{minted}

\usepackage[group-separator={,},group-minimum-digits=4]{siunitx} 

\definecolor{dkgreen}{rgb}{0,0.6,0}
\definecolor{gray}{rgb}{0.5,0.5,0.5}
\definecolor{mauve}{rgb}{0.58,0,0.82}
\definecolor{lightgray}{gray}{0.90}
\definecolor{ms-office-blue}{rgb}{0.36, 0.60, 0.83 }

\newcommand{\projectname}{{SSIoT}\xspace} 
\newcommand{\ssiot}{\projectname}
\newcommand{\SSIoT}{\projectname}
\newcommand{\DIYL}{\projectname}

\newcommand{\CMU}{Carnegie Mellon University}

\newcommand{\encryptedDataKey}{$\mathrm{Enc}[\mathrm{Enc}(K)_{K_A}]_{K_M}$\xspace}
\newcommand{\encryptedDataFile}{$\mathrm{Enc}[\mathrm{Enc}(data)_{K_A}]_{K}$\xspace}

\usepackage{enumitem}
\setitemize{noitemsep,topsep=0pt,parsep=5pt,partopsep=5pt}

\renewcommand{\paragraph}[1]{\vspace{2pt}\noindent\textbf{#1}}

\begin{document}

\date{}

\title{Self-Serviced IoT: Practical and Private IoT Computation Offloading with Full User Control}

\author{
{\rm Dohyun Kim}\\
\CMU
\and
{\rm Prasoon Patidar}\\
\CMU
\and
{\rm Han Zhang}\\
\CMU 
\and
{\rm Abhijith Anilkumar}\\
\CMU
\and
{\rm Yuvraj Agarwal}\\
\CMU
} 

\maketitle

\begin{abstract}
The rapid increase in the adoption of Internet-of-Things (IoT) devices raises critical privacy concerns as these devices can access a variety of sensitive data. The current status quo of relying on manufacturers' cloud services to process this data is especially problematic since users cede control once their data leaves their home. Multiple recent incidents further call into question if vendors can indeed be trusted with users' data. 

At the same time, users desire compelling features supported by IoT devices and ML based cloud inferences which compels them to subscribe to manufacturer-managed cloud services. An alternative to use a local in-home hub requires substantial hardware investment, management, and scalability limitations.  
This paper proposes Self-Serviced IoT (\projectname), a clean-slate approach of using a hybrid hub-cloud setup to enable privacy-aware computation offload for IoT applications. Uniquely, \projectname enables opportunistic computation offload to public cloud providers (e.g. Amazon AWS) while still ensuring that the end-user retains complete end-to-end control of their private data reducing the trust required from public cloud providers. We show that \projectname can leverage emerging function-as-a-service computation (e.g. AWS Lambda) to make these offloads cost-efficient, scalable and high performance as long as key limitations of being stateless, limited resources, and security isolation can be addressed.  
We build an end-to-end prototype of \projectname and evaluate it using several micro-benchmarks and example applications representing real-world IoT use cases. Our results show that \projectname is highly scalable, as compared to local-only approaches which struggle with as little as 2--4 apps in parallel. We also show that \projectname  is cost-efficient (operating a smart doorbell for \$10 a year) at the cost of minimal additional latency as compared to a local-only hub, even with a hardware ML accelerator. 
\end{abstract}

\section{Introduction}

The adoption of Internet of Things (IoT) devices such as voice assistants, door locks, and smart cameras has been increasing\cite{Gartner-projection}. These devices often access a variety of personal data, such as audio and images/video to support ``smart'' functions (e.g., notifying a homeowner with a video clip of a visitor \cite{ring-doorbell} or if someone other than the occupant is detected \cite{nest-familiar-face}). To support these functions, manufacturers often bundle their hardware with proprietary cloud services, either as subscriptions or add-on services (e.g. Nest Aware or Ring Protect Plans from \$6 -- \$20 a month), sending sensitive user data to their cloud for data storage, processing, and ML-based inferences.  
This model of relying on the manufacturers' cloud services to process this data raises serious privacy concerns for end users since they cede control once their data leave the confines of their home on how their data is used and who it is shared with. 
While privacy laws (e.g. GDPR and CCPA) help hold manufacturers accountable, their enforcement is limited, and recent incidents show that even reputable companies are prone to privacy breaches (e.g., employees of the company or even other users gaining unauthorized access to users' data~\cite{ring-unfettered-access:2019, swann-wrong-user:2018, alexa-listening:2018}). Recent user studies confirm that users have serious privacy concerns around IoT devices~\cite{eig:iot_report:2018} and want more transparency and control over the use of sensitive data from these devices ~\cite{Emami-Naeini:privacy-iotpurchase:2019,iot_chasm:2016}.

To mitigate privacy concerns from IoT devices sending data to the manufacturer's cloud, one approach is to keep all data confined to an in-home hub and run applications locally. Several open-source IoT platforms (e.g., OpenHAB \cite{openhab}, Home Assistant \cite{home-assistant}) enable this \emph{local-only} model for IoT devices supporting simple home automation tasks. However, this local-only approach is limited by the local hub's computation capability. Low-cost hardware platforms (e.g. Rasberry Pi) cannot support rich IoT applications that often use complex ML models, requiring significant upfront investment in hardware accelerators~\cite{EdgeTPU, JetsonNano} or computers with GPUs. In fact we show that inferences using common ML models for tasks such as object or face detection, can take between 99ms -- 4.3s \textit{per inference} on current generation hardware that can serve as a local-hub. The hardware requirement only becomes worse as ML models become more complex, or as IoT devices are added to a home, rendering a hardware hub inadequate in a few years adding to maintenance and replacement costs. A natural extension to augment the computation at a local hub is to offload heavyweight computation, like ML-based inferences, to the cloud and numerous approaches to do this exist ~\cite{aws-greengrass,azure-iot,Noghabi:Steel:2018,SmartThings, Beam:Shen:2016,Beam:Singh:2015,EdgeABC:2020}. However, the key limitation with these generic solutions is that they do not provide mechanisms to ensure that the user retains complete end-to-end control of the privacy of their IoT data and the computations performed on it. 

In this paper, we present \SSIoT (Self-Serviced IoT), a novel approach that addresses the limitations with prior offloading approaches and provides privacy-aware computation offload in a hybrid hub-cloud setting for IoT use cases. A unique insight in \projectname is that emerging general-purpose function-as-a-service (FaaS) platforms, such as AWS lambda, provide the right abstraction to offload computation for IoT workloads. Specifically, FaaS decompose applications into `triggers' (events) and `actions' (functions) and manage all the runtime aspects of executing these functions~\cite{lynn2017preliminary}. FaaS platforms are stateless, and the cost model is ``pay per invocation'' thereby incurring no `idle' costs as compared to having dedicated VMs. They are also highly scalable to allow essentially a huge number of invocations in parallel. As a result, FaaS platforms are uniquely positioned as a cost-effective option for event-driven smart home and IoT  applications, where events of interest are often correlated in time and bursty with long periods of no activity (e.g. person is asleep, or the home is unoccupied)~\cite{oconnor2019homesnitch}. However, a key challenge with using FaaS platforms is data privacy and security, to ensure that input data to invoke the cloud function and the response are end-to-end private as well as preventing any data exfiltration from the FaaS platform itself. Another key challenge with using FaaS is to decide when to offload to the cloud and ensure that we meet IoT application requirements. To address these challenges, we design and implement our end-to-end \SSIoT framework, which includes deployment toolchains, creating isolated runtimes, and key-management services (KMS), that collectively enable users to ``self-service'' complex IoT applications using their own FaaS cloud account. Ultimately, \SSIoT allows users to be in complete control of how their own IoT apps collect, process, and share their sensitive data.

We build and evaluate a prototype of \SSIoT using a suite of workloads representing emerging IoT applications (e.g. image classification and object detection). We compare \SSIoT's offloading performance by running apps on two types of local devices: Raspberry Pi (\$30-\$35) and Jetson Nano ($>$\$100, with hardware accelerators). \SSIoT can reduce latency up to 80\% compared to just using a local-only Raspberry PI device. While a Jetson Nano with an accelerator has better performance (31\%-81\% lower latency), \SSIoT offers significant benefits when scaling to multiple concurrent applications as compared to the Jetson, which can only serve 2 application instances. Moreover, \SSIoT is cost-efficient: processing over 98k machine learning requests for \$1 and operating real-world applications like a smart doorbell for a year for less than \$10. Moreover, our opportunistic offloading strategy further reduces the operating costs by 15\%-20\% by utilizing local resources when available. Our prototype implementation is based around an existing open-source IoT software hub (OpenHAB), providing a path to adoption. For example, our smart doorbell application implemented on \projectname (similar to the Ring Doorbell) comprises only 18 lines of code.

In summary, we make the following contributions:

\begin{itemize}
	\item We propose \SSIoT (Self-Serviced IoT), an IoT computation offloading framework for apps to process users' private data. We leverage FaaS platforms to augment the limited computing power of a local hub while preserving users' control via the design of \SSIoT Key Management Service.

	\item We build an end-to-end prototype of \SSIoT using off-the-shelf hardware (Raspberry Pi and Jetson Nano) and cloud services (AWS Lambda). We address several practical challenges to reduce cost, latency, and performance overheads while facilitating developer adoption.

	\item We evaluate our \SSIoT prototype using benchmarks to show that \SSIoT is scalable ($>16$ concurrent apps) and cost-effective (serving up to 98k requests for \$1). Our end-to-end integrated smart doorbell app built using \projectname provides similar functionality as commercial devices while protecting the users' data privacy and costing just \$10 a year.

\end{itemize}

\section{Motivation}
\label{section:background}
We motivate below the need for \projectname by highlighting a number of privacy breaches related to IoT devices and their data, and why existing approaches are insufficient in terms of protecting user privacy. 

\subsection{Privacy Incidents and User Preferences}
\label{sec:motivation-incidents}

Currently, users need to fully trust manufacture-managed app services running on the cloud to use many of the smart features supported by IoT devices. In this section, we examine several real-world incidents where manufacture-managed app services compromise end-users' privacy, grouped into three categories.

\paragraph{- Internal Company Privacy Policy.} A recent report \cite{ring-unfettered-access:2019} indicated that employees of a smart doorlock vendor had unlimited access to both live and archival video recordings from user devices without any access control protection, along with users' email addresses. This incident shows that privacy violations can happen without users noticing them. Users cannot monitor or audit how the vendor handles their data in the cloud. Anyone could potentially have unfettered access to their private data.

\paragraph{- Human Errors.} IoT devices that collect users' sensitive video \cite{swann-wrong-user:2018} and audio \cite{alexa-listening:2018} allowed one customer to access another user's private archive of audio and video due to human error.
Although the specifics of the incidents are unclear since those incidents happened inside the vendors' backend, it provides evidence that even minor mistakes on the backend can potentially lead to serious privacy breaches.

\paragraph{- Compromised Central Services.} Since IoT app services store users' data in a centralized database, a single breach can lead to exposure of data about all users and devices. For example, an attacker can spoof an IoT device's serial number to look up other device's status in the backend cloud \cite{swann-vulnerability:2018} or get precise user locations by intercepting the plaintext network packets between devices and the cloud app service \cite{google-home-location-leak:2018}.

\paragraph{User Demands for Privacy.}
These IoT privacy threats demonstrate that manufacturer-managed app services can lead to severe privacy breaches and suggest that decentralizing the functionalities of these app services is a key for protecting users' privacy. Recent studies show that users indeed want more control over their private data collected by IoT devices. In one study, 92\% of 1629 global participants wanted control over their data and demanded transparency about automatic data collection \cite{eig:iot_report:2018}. In a separate study, an overwhelming majority of participants wanted to be notified of data collection practices at the time of IoT device purchase \cite{Emami-Naeini:privacy-iotpurchase:2019}. 

\subsection{Limitations of Existing Approaches}

One obvious approach to address users' privacy concerns is to restrict all IoT apps to process user data and perform computation locally, as demonstrated by the growing popularity of self-hosted automation platforms such as OpenHAB~\cite{openhab} and Home Assistant~\cite{home-assistant-analytics}.  This \emph{local-only} approach prevents any data from leaving users' homes but requires users to install powerful local machines to handle apps' computation demands. As users install more IoT devices in their homes and adopt more complex smart applications (e.g., using larger deep learning models for inference), they have to upgrade to even more capable hardware or install computation accelerators (e.g., GPUs, Coral~\cite{coral-accelerator}) on a regular basis, incurring maintenance overheads in terms of time, money, and expertise.

To augment local machines' capability, users can offload computation by leasing virtual machines from cloud service providers. This saves the upfront cost of investing in local machines and provides an easy path for upgrades. Unfortunately, this approach is rather wasteful (i.e., users need to pay for the machine's idle time) and still suffers limitations for IoT use. In a home setting, interesting events often happen in a bursty manner~\cite{oconnor2019homesnitch}, where multiple devices/smart apps compete for shared computing resources in a short period. For example, an intruder alert system may simultaneously access and coordinate multiple devices (e.g., motion sensors, cameras, doorbells) in a home to detect the presence of strangers. As we will show in our evaluation, it is challenging for a single machine to perform concurrent heavy-weight computations (e.g., deep learning inferences) with acceptable latency guarantees, even if they are optimized for edge inferences (such as Jetson Nano~\cite{JetsonNano}).
To address bursty workloads, users need to have multiple machine standby --- an expensive option --- or to look for more scalable and cost-efficient alternatives (as we propose in \SSIoT).

\section{\SSIoT Overview}
\label{sec:system_overview}

In this section, we start with summarizing our design goals, articulate design choices for \ssiot (\S \ref{sec:objectives}), and describe our threat model (\S \ref{sec:threat_model}). Next, we discuss target users and the type of applications supported by \ssiot (\S \ref{subsection:target_users_usecases}) and provide a high-level overview of \ssiot architecture (\S \ref{subsection:system-architecture}). Next, we present the design of \ssiot Hub, which helps streamline deployment and execution of \ssiot applications (\S \ref{subsection:ssiot_hub}), and \SSIoT Key Management Service, which protects users' privacy in the case of cloud offloading (\S \ref{sec:encryption-design}).

\subsection{Design Goals}
\label{sec:objectives}

Our primary motivation is to protect user's private IoT data, by running applications on the local hub, or opportunistically offloading it to the cloud in a cost efficient and privacy preserving manner. 

\paragraph{Cost Effectiveness and Scalability.}
To minimize the cost of running \SSIoT applications, we leverage off-the-shelf \emph{function-as-a-service (FaaS)} -- also known as \emph{serverless} -- platforms, a popular offering from several cloud providers (e.g., Amazon Lambda \cite{aws-lambda}, Google CloudFunction \cite{google-gcf}).
FaaS platforms allow users to provide a custom, stateless function to run on the cloud. Then, the service provider takes care of all the management tasks such as runtime management, load balancing, and scaling.
Notably, these functions are \emph{stateless} so that each invocation is self-contained and independent. Thus, a key advantage of the FaaS platform is its unique cost model: users are only charged per function invocation and how long the function execution takes in terms of CPU time, rather than having to pay for any idle machine time. In contrast, spinning up a Virtual Machine on a cloud provider incurs cost whether or not the VM is busy or is idle.   
Therefore, by offloading functions, as needed to FaaS platforms, \SSIoT eliminates the need for expensive and powerful local hubs to guarantee good application performances. \SSIoT's cost and performance advantages stand out as more IoT apps execute concurrently due to the bursty nature of many IoT scenarios~\cite{oconnor2019homesnitch}.

\paragraph{Preserving End Users' Privacy.}
While FaaS platforms provide isolation among function runtimes, we also need to address two key design challenges in \SSIoT: ensuring that users' private data is \emph{only accessible} by authorized application runtime and preventing \emph{malicious applications} from leaking users' private data. To prevent unauthorized access to private data, we need to make sure that no one except the intended application runtime can decrypt raw data. We designed a lightweight key management service (KMS) to facilitate confidential communication between the user's local \SSIoT hub and the remote application instance (\S \ref{sec:encryption-design}). 
Users can host their own KMS (e.g., on their local hub or a separate server) or use KMS's offered by independent, trusted entities.
To protect users from untrusted application code provided by other developers, we propose an enforcement mechanism for remote execution of untrusted application code in the cloud.

\paragraph{Low Latency and High Performance.}
We also aim to optimize the latency of running applications, specifically addressing the new challenges arising from the use of FaaS offloading. One common issue with FaaS platforms is their cold-start latency. Active functions are considered ``warm'' instances, and they will enter a ``cold'' state after receiving no invocation for extended periods of time. Cold functions take a significantly longer time to invoke, but all subsequent calls will be served at ``warm'' instances~\cite{peeking_serverless:2018}.
To mitigate the impacts of cold-start latency, users can instruct the hub to periodically send keep-alive requests to keep their apps in a warm state on the FaaS platform. 
Our evaluation shows that doing so increases operational costs slightly.
In addition, \SSIoT supports an alternative opportunistic offloading strategy that balances costs and performance. 
The local hub can keep serving apps only offload to FaaS platforms when it lacks sufficient resources to process new app instances.

\subsection{Threat Model}
\label{sec:threat_model}

We assume an adversary who could intercept and manipulate network traffic in an attempt to access private data. 
We assume local smart devices may be malicious and attempt to leak user data over the Internet. Therefore, we do not allow these devices to connect to the Internet directly. To achieve this, many prior works have proposed effective network isolation solutions~\cite{zhang2021capture, hong2018bark, erickson2018dreamcatcher, trimananda2018vigilia}. 
Therefore, we inherently assume no Internet connectivity for \SSIoT-enabled devices throughout our design. 
Similarly, we assume the smart applications, subject to cloud offloading, may be malicious as well. Therefore, we need to enforce runtime isolation to prevent data leaks.

We consider the cloud service providers as honest but curious entities. We assume they will not actively extract user data inside the running containers and VM memory regions, as indicated by common industry practices and providers' end user agreements \cite{aws-data-privacy}.
Meanwhile, the FaaS platform and other co-located FaaS applications may potentially obtain network traffic, as mentioned in our attacker models. Our goal is to prevent unauthorized access to users' private data in such scenarios. While it would provide an even stronger security guarantee to assume an \emph{active} attacker model for cloud service providers, we leave such consideration for untrusted public cloud providers to related and future work (e.g., prototypes for processing private data within secure enclaves \cite{SeLambda}, enabling private machine learning inferences with CPU/GPU secure enclaves \cite{Slalom, volos2018graviton}). Finally, we do not consider vulnerabilities orthogonal to our design goals, such as social engineering and weak credentials in our threat model.

\subsection{Target Users and Use Cases}
\label{subsection:target_users_usecases}

\paragraph{Target Users.} 
\SSIoT is targeted towards users who prefer to avoid manufacturer-provided cloud apps due to their lack of transparency. 
In our prototype, we extend the open-source home automation hub's (OpenHAB) web dashboard to provide \SSIoT device integration.
To use \SSIoT, users need to learn how to initialize new devices to an \SSIoT hub in their local environment and be familiar with configuring IFTTT-style trigger-action rules. We acknowledge that these requirements make deploying \SSIoT harder for normal users, who are not as tech-savvy as the home automation enthusiast community. Going forward, we plan to extend \SSIoT's UI to simplify the new application setup process, improving the system's usability and lowering the barrier to adoption. 

\paragraph{Target Applications.} 
\SSIoT is designed for event-driven smart home devices and automation applications (e.g., push notification if a doorbell detects someone on the doorstep). These use cases can tolerate a reasonable amount of latency since typical human interactions can take several seconds.
As a reference point, existing home automation applets on platforms like IFTTT already take several seconds to process requests~\cite{zhang2021capture, mi2017ifttt}. 
We acknowledge \SSIoT may not be suitable for delay-sensitive applications, but the sheer number of cloud automation applets~\cite{mi2017ifttt} show that \SSIoT can help with many common deployment scenarios to protect private data.

\begin{figure*}[t]
    \centering
    \includegraphics[width=0.9\textwidth]{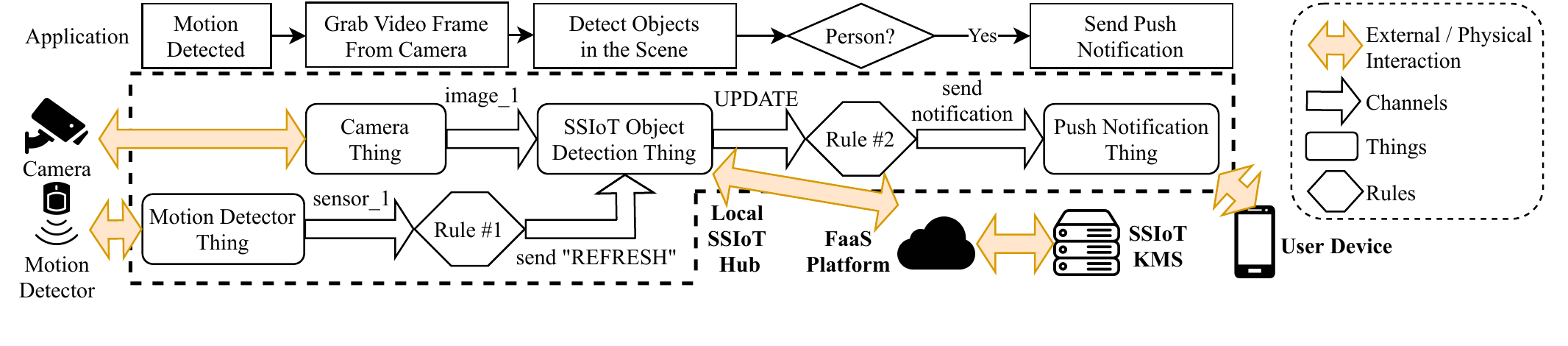}
    \vspace{-3mm}
    \caption{\SSIoT architecture. The top half explains the example application's workflow. The rules are defined in \Cref{fig:rule_example}.
    }
    \label{fig:application_model} 
\end{figure*}

\begin{figure}[t]
\begin{minted}[linenos,mathescape,numbersep=5pt,xleftmargin=10pt,frame=lines,fontsize=\footnotesize]{verilog}
rule "Rule#1"
when
    Thing "sensor_1" changed from "off" to "on"
then
    ssiot_object_detection.sendCommand("REFRESH")
end

rule "Rule#2"
when
    Item "ssiot_object_detection" received update
then
    if label == "person" && score > 0.80 {
        sendNotification("detected a person")
    }
end
\end{minted}
    \caption{User-defined rule examples. Rule \#1 states that the motion sensor will trigger the object detection pipeline when a movement is detected. 
    Once the offloaded object detection function returns, Rule \#2 checks whether there is any person detected and sends a notification to the user's device.}
    \label{fig:rule_example}
\end{figure}

\subsection{System Architecture}
\label{subsection:application_model}
\label{subsection:system-architecture}

\Cref{fig:application_model} illustrates the high-level architecture of \SSIoT with an example  application including a camera and a motion sensor. The camera alerts the user when it detects a person in front of the door. 
User needs to install a \SSIoT hub in their home and set up all local devices with the hub. In this example, we show two drivers -- Camera Thing and Motion Detector Thing -- that connect the devices to the hub. 
We denote different programs running on the hub as ``Things'', such as device drivers and implementation of functionalities (e.g., detecting objects and sending notifications).
In addition, applications are developed with ``Rules'' -- code snippets expressed using   if-this-then-that (IFTT) rules. 
\Cref{fig:rule_example} lists the two examples in the smart doorbell. These rules specify how to connect different things together and branching logic.

In this example, the \SSIoT object detection thing offloads computation to the remote cloud. \SSIoT leverages function-as-a-service (FaaS) platforms by packaging certain functionality (e.g., machine learning inference based on input image) as stateless, FaaS apps. The hub sends user's private data to the FaaS app instance for function execution. To access data, the FaaS app needs to contact an independent \SSIoT Key Management Service (KMS) to decrypt the content. Users can host their own \SSIoT KMS or use trusted third parties. We explain the KMS functionality further in \Cref{sec:encryption-design}.

\begin{figure}[t]
\centering
\includegraphics[width=0.99\columnwidth]{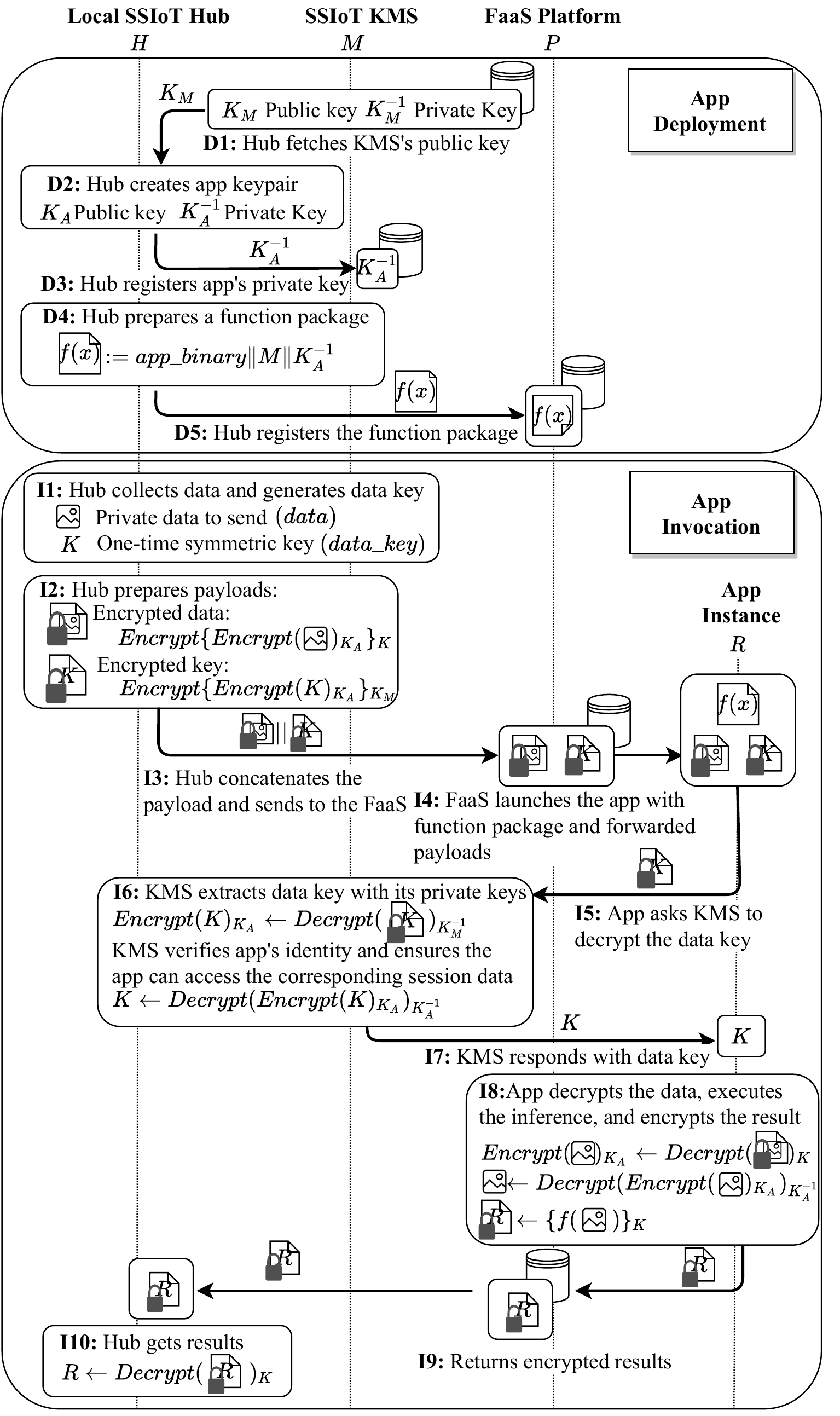}
\caption{Procedure for deploying and invoking an application. \Cref{subsection:system-architecture} provides a high-level overview of app deployment and invocation processes, while a detailed protocol walk-through can be found at \Cref{sec:appendix-workflow}.}
\label{fig:procedure} 
\vspace{-3mm}
\end{figure}

\Cref{fig:procedure} shows the detailed procedure for single application deployment and invocation with \ssiot Hub. For deploying an \ssiot application, the \ssiot hub registers the application with KMS and deploys a packaged application to the FaaS platform (\textbf{D1--D5}). For every invocation, the \ssiot hub prepares the payload using raw input data and sends requests to the FaaS platform to launch a remote application instance (\textbf{I1--I4}). Upon invocation, the application instance contacts the KMS to get the key to decrypt the data payload (more details in \S \ref{sec:encryption-design}) execute the function, and return results to the \ssiot hub (\textbf{I5--I10}). We provide further details in \Cref{sec:appendix-workflow}.

\subsection{SSIoT Hub}
\label{subsection:ssiot_hub}

We now describe the various components of the \ssiot Hub, which enables application deployment and invocation efficiently with minimal user intervention.

\paragraph{OpenHAB Integration.} Our application model and terminologies are based on the design of OpenHAB~\cite{openhab}.  OpenHAB provides a user-friendly web interface to navigate users through device management and application creation processes. \SSIoT extends OpenHAB's codebase with additional binding implementations, such as \SSIoT Object Detection, to enrich application functionalities. In the future, we envision third-party developers can provide additional implementations (open- or closed-sourced) for new bindings and holistic applications tailored for a variety of tasks and machine learning models. We are optimistic about this vision since other communities already embrace a similar model (e.g., IFTTT user-shared applets~\cite{ifttt:2019} and Docker hubs~\cite{docker-hub}).

\paragraph{\SSIoT Toolchain and Runtime.}
To help the local \SSIoT hub with deploying new apps and managing running instances, we design and implement several \SSIoT programs, including the Toolchain, Local Runtime, and Remote Runtime.
The \SSIoT Toolchain is a fully trusted software module for the local hub. It handles tasks like app key pair generation and app binary packaging for deployment. 
The local \SSIoT runtime interfaces with local devices during app invocation. It generates fresh data keys and performs all cryptography operations related to invocation.
Meanwhile, the remote \SSIoT runtime performs similar functions for the FaaS app. All of these programs help streamline new app creation and execution, reducing the burden of end-users and app developers. 

\paragraph{Opportunistic Offloading.}
Both local-only computation and FaaS offloading have their unique advantages. We propose a new opportunistic offloading strategy to combine the best of both approaches. If the local device has sufficient computation resources (especially for embedded accelerators like the Jetson Nano), executing functions locally can often have lower latency and operational costs (e.g., the cost of electricity to power it). On the other hand, offloading to a FaaS cloud significantly augments the limited resources of local devices, improving scalability and performance under heavy workloads.

Therefore, we add an opportunistic offloading strategy to \SSIoT that balances the costs incurred for running cloud functions while providing high performance and low latency by leveraging local compute resources. To support opportunistic offload, an \SSIoT hub includes three major components and manages two service queues. The first component is the input data management process, which polls for data from IoT devices and forwards it to a data queue for processing. The second component is a resource allocator that manages the \SSIoT hub's local computation resources. The resource allocator supports user-customizable policies, such as minimizing overall latency, maintaining cloud cost budgets, or a balance between cost and latency. We assumed the default goal is to provide the best performance for this example.
If the local device has sufficient compute available, the resource allocator redirects the request from the data queue to local serving instances and updates the resource monitor; otherwise, it will forward the request to the FaaS cloud app instance. 
In either case, the computed results are pushed to a results queue. Next, the third component - a result-handling process reads the results from this queue and sends them back to the caller (i.e., automation apps and bindings on the hub).

\subsection{Encrypting Private Data and Isolation}
\label{sec:encryption-design}

To protect users' privacy as stated in our design goals (\S \ref{sec:objectives}), one potential approach is to have the local hub negotiate and derive unique session keys with the app instance running on the FaaS. Doing so introduces additional delays in key negotiations for every function invocation, and the process unnecessarily repeats if the user wants to execute multiple apps for the same data in parallel. We also cannot package the unique public-private key pair into the FaaS app binary since the platform reuses the same binary to create multiple app instances (which should each have unique keys).

To overcome these challenges, we propose the design of \SSIoT Key Management Services (KMS). The KMS is a trusted entity and users can deploy their own KMS instances on the local \SSIoT hub, standalone machines, or utilize any other trusted service (e.g., to be integrated into secure key management services from public cloud providers~\cite{aws-kms, azure-key-vault}). Deploying KMS in a user's cloud alongside the FaaS deployment reduces key decryption latency, whereas deployment of KMS on the local hub provides complete user control with some increase in communication overhead.

The KMS manages keys for various applications and assists with data decryption at a per-request level. As shown in \Cref{fig:procedure}, the hub generates a fresh data key ($K$) for each app input data file (e.g., an image or audio recordings). It then constructs two encrypted messages, encrypted data content (\encryptedDataFile) and encrypted data key (\encryptedDataKey), to share with the remote app instance. 

The app cannot decrypt the data simply based on these two messages. It needs to contact the KMS to retrieve the actual data key. 
Since the KMS server is trusted, this design enables users to gain real-time visibility to see which app accesses their data and improves accountability. 

To access user data, the app instance needs to forward the encrypted data key to the KMS. The KMS enforces access control policies for the app instances on behalf of the user. Users can obtain real-time insights on how their data is accessed, who have accessed it, and revoke the app's permission. If the app instance is capable of accessing the data, the KMS then extracts the data key (\textbf{I6} of \Cref{fig:procedure}) and replies the key to the app. The actual data key is sent over a TLS session, so the FaaS platform cannot intercept the key. With the data key and its app private key, the app can now access users' data and execute functions on it (\textbf{I8}). 

Another important requirement for \DIYL runtime is to isolate the remote IoT app from the FaaS platform's network and file system. We have extensively explored the design space of the \DIYL runtime and applications that we can support. We leverage runtime enforcement and sandboxing since statically analyzing generic applications can be computationally challenging. In our current design, the app can be either a pre-trained DNN or Java's JAR package. We achieve network and file system isolation \emph{at compile time} for DNN applications. To support generic, imperative types of applications, we use the Java Virtual Machine (JVM), which provides primitives to isolate the applications' access to the network and the filesystem \emph{at runtime}. We also considered leveraging either kernel or container-level security primitives such as \texttt{seccomp} syscall or configure process namespaces (i.e., creating a new child namespace under the isolated environment's namespace), but these methods require higher privileges on the host system that is beyond the current settings provided to cloud functions.

\section{Implementation}
\label{section:implementation}

We implemented a prototype of \SSIoT using the popular AWS Lambda FaaS platform \cite{aws-lambda}, and describe the various components below. 

\paragraph{\SSIoT Runtime \& Toolchain.}
The remote and local \SSIoT runtimes are implemented in 1562 lines of Java 1.8, one of the officially supported runtimes for AWS Lambda. For cryptographic operations, we make use of open-source \texttt{bouncycastle}'s \texttt{openpgp} library \cite{bouncycastle}.
\SSIoT toolchain contains generic C++ code that binds deep learning applications (compiled by TVM~\cite{tvm:2019}) and serializes data communications in \texttt{flatbuffers} \cite{flatbuffers}.
Our \SSIoT toolchain also contains a set of scripts to automate compiling a pre-defined DNN model into a binary and merging the compiled binary and the \SSIoT runtime.
To support generic, imperative types of applications, we use the Java Virtual Machine (JVM), which provides primitives to isolate the applications' access to the network and the filesystem \emph{at runtime}.

\paragraph{Packaging Apps.}
A deployable package has two main components: an application binary and remote \DIYL runtime. The application binary is the actual code a user wants to run remotely, while remote \DIYL runtime includes \DIYL-specific application logic that handles the app invocation procedure. When an app is invoked, the \DIYL runtime decrypts the payload, spawns a separate process for the application binary, and passes the plaintext data to the application binary via standard input.
Currently, users can choose one of MXNet Gluon's vision model zoo~\cite{gluon} for image classification and object detection. Given the model name, our \SSIoT toolchain downloads the network configuration and parameters from the model zoo. Then, it starts a pre-built TVM docker image and compiles the network into an \texttt{x86\_64} binary.

\paragraph{\SSIoT KMS.}
We prototyped our KMS in 372 lines of Go, extending \texttt{golang}'s \texttt{openpgp} implementation \cite{golang-openpgp-impl}. The service is built as a webservice with the three REST APIs over HTTPS (TLS v1.2 enabled). \texttt{GetKMSPublicKey()} returns the public key of the KMS. \texttt{RegisterAppPublicKey($K_{A}$)} register the app's public key to the KMS that allows the KMS to decrypt the encrypted key blob. \texttt{DecryptDataKey(encrypted\_data\_key))} decrypts the encrypted key blob received by function instance.

\paragraph{IoT Hub Integration.}
We implement the \SSIoT hub integration as a binding of OpenHAB \cite{openhab}, one of the open-source implementations of IoT hub, in 608 lines of Java code. Instead of building a completely new IoT hub stack, we decided to leverage the rich connectivity, IoT device drivers, and inter-operability of OpenHAB. OpenHAB supports key functionalities such as rule engines, event channels, and a set of user interfaces (UI) for end users to configure our \SSIoT addon. Other than the binding addon implementation, no code change was required to integrate \SSIoT to OpenHAB. We believe this is a significant advantage of \SSIoT since it can also be similarly integrated with other extensible IoT hub architectures such as HomeAssistant~\cite{home-assistant} or any others that become popular in the future.

\section{Evaluation}
\label{sec:evaluation}

Our evaluation aims to answer the following questions.

\begin{itemize}
    \item How does \SSIoT's cloud offloading affect the overall latency of the application, especially when compared to local-only alternatives? 
    \item How does the cold start latency, a common challenge for FaaS platforms and offloading, impact \SSIoT? Can \SSIoT's mitigation strategy and opportunistic offloading design alleviate this latency effectively? 
    \item What are \SSIoT's unique advantages and benefits over local-only designs? Can \SSIoT achieve its practicality goal while offering privacy protection? Specifically, is \SSIoT scalable and cost-effective? 
    \item Can our current \SSIoT prototype be used to express real-world IoT applications to show that \SSIoT is realistic?
\end{itemize}

To measure the performance of \SSIoT, we integrated a separate Raspberry Pi that captures camera images as a local smart device into \SSIoT. We develop a suite of benchmark applications that perform inference tasks on these images. 
We evaluate the application executions on different hardware platforms serving as \SSIoT Hubs, including low-cost single-board computers (Raspberry Pi 3, \$30) and hardware AI accelerators (Nvidia Jetson Nano, \$100).
We also characterize the computational complexity of each \SSIoT component (e.g., data encryption, cloud communication, remote execution)  in terms of execution time to understand the limiting factors to scalability. Our results show that the major source of \SSIoT latency is the communication overhead and the lack of hardware accelerators (CPU-only, no GPUs). Despite these overheads, cloud offloading provides significant benefits in terms of scaling to multiple concurrent apps and reducing operating costs.

\subsection{Latency Comparison}
\label{sec:eval-latency}

We compare the function execution latency between \SSIoT's FaaS offloading and local-only approaches. Our results show that FaaS offloading outperforms cheap single-board computers (Raspberry Pi) because the cloud FaaS runtime has more capable hardware. On the other hand, FaaS offloading's network communication overhead and lack of GPU support cause it to perform slower than relying on local hardware accelerators (e.g., Jetson Nano) by noticeable margins (5x slowdown).
In addition, the cold start latency has a significant impact on FaaS executions, but we show that our proposed mitigation techniques address this challenge.

\begin{figure}
    \centering
    \includegraphics[width=0.98\columnwidth]{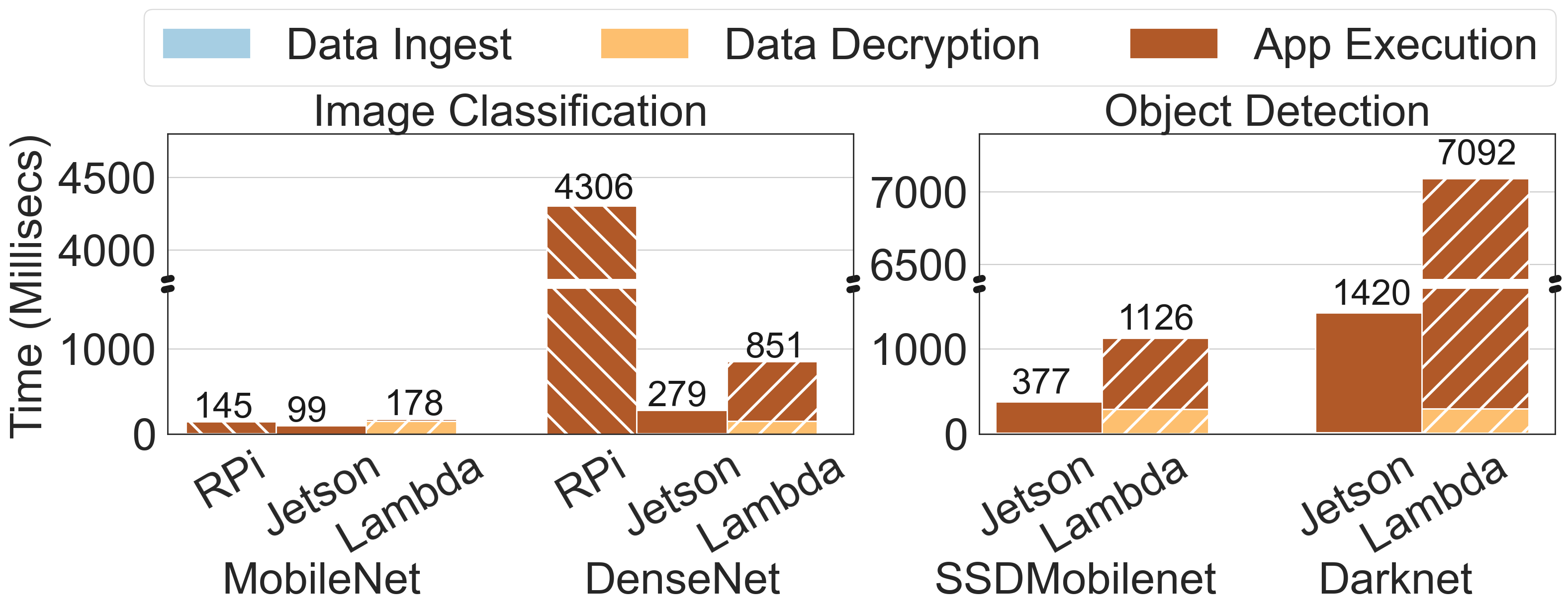}
    \caption{Average inference latency across different platforms and models. 
    Raspberry Pis cannot perform object detection due to limited resources.
    Moving apps to lambda reduce latency for single board computers (RPi), but the network communication overhead overshadows the processing benefit when compared to hardware AI accelerators (Jetson Nano).
    }
    \label{fig:eval-function-latency}
\end{figure}

\paragraph{Effect on Latency due to Hardware Capabilities.}
\Cref{fig:eval-function-latency} compares the average latency for image classification and object detection tasks executed 100 times each on a Raspberry Pi 3 Model B+, Jetson Nano, and AWS Lambda (in warm states).
Due to space constraints, we include results for more models in \Cref{sec:additional-eval-results}.
The compute capability of Raspberry Pi is fairly limited. It cannot execute any of the object detection models, and it executes image classification apps at a relatively slow speed. 
Offloading these apps to Lambda instances reduces latency by as much as 80\% in the case of DenseNet. As for really small models like MobileNet, Lambda performs slightly worse due to the additional cloud communication overhead.

Meanwhile, Jetson Nano, an embedded system with AI hardware accelerators, outperforms both Raspberry Pi and Lambda, reducing latency between 31\%-81\% from the fastest alternatives.
Although it seems that Lambda is unappealing when compared to Jetson Nano, we want to point out that these latency overheads should still be acceptable. Previous studies find typical home automation applications often have several seconds of latency, mostly caused by delays in the backend cloud services~\cite{zhang2021capture,mi2017ifttt}. 
\SSIoT incurs $<$1 second for image classification tasks and only a few seconds for more complicated object detection models.
Therefore, in spite of noticeable latency overhead, \SSIoT is still a competitive option for its advantages we will show in later sections.

\begin{table}[t]
    \centering

    \small
    \begin{tabular}{c|c|c}
    \hline
        \textbf{Latency} & \textbf{Image Classification} & \textbf{Object Detection} \\
        \textbf{Comparison} & \textbf{(DenseNet)} & \textbf{(Darknet)} \\
    \hline \hline
    \textbf{Cold Instance} & 9153 ms & 35959 ms \\
    \hline
    \textbf{Warm Instance} & 851 ms & 7092 ms \\
    \hline
    \end{tabular}
	\caption{Average inference latency (in milliseconds) between cold and warm AWS lambda instances.
    Cold states are much slower ($>$11x) due to delays in app instantiation.}
	\label{tab:eval-function-coldwarm-latency}
\end{table}

\paragraph{Cold Start Latency.}
A common issue with FaaS platforms is the cold start latency, which also affects \SSIoT's overall performance. 
\Cref{tab:eval-function-coldwarm-latency} compares the average latency between invoking apps in warm and cold states. 
Cold-start latency incurs significant overhead (5x-36x) due to app instantiation and fetching function implementations from FaaS platform backends. 
Even if the app is in a cold state, the cold-start latency only affects the very first incoming requests after an extended idle period. All subsequent requests will reach the ``warm'' app instances and execute quicker.

We propose a number of mitigation strategies to avoid this cold-start latency.
The most straightforward solution is to periodically send requests to keep FaaS instances stay in the ``warm'' state.  AWS Lambda users can use serverless warm-up plugins \cite{lambda-warmup-plugin} to do so.
These periodical messages increase the app invocation frequency and hence add cost.
We conduct an experiment to ensure this periodical keep-alive message approach does not become cost-prohibitive. We send keep-alive requests every 15 minutes, well below the observed 26-minute Lambda idle threshold~\cite{peeking_serverless:2018}.
Our results confirm that all of our requests are served by the ``warm'' instances. This approach introduces nominal additional costs. As we will show in \Cref{subsubsection:cost_analysis}, sending requests once every 15 minutes usually cost way less than \$1 \emph{per month} (2,880 requests), depending on the model sizes. Another alternative is to use our opportunistic offloading strategy, which can utilize local computation to address requests till Lambda instances warm up, thus reducing the operational costs at the expense of higher processing demand from the hub.

\begin{figure}[t]
    \centering
    \includegraphics[width=0.98\columnwidth]{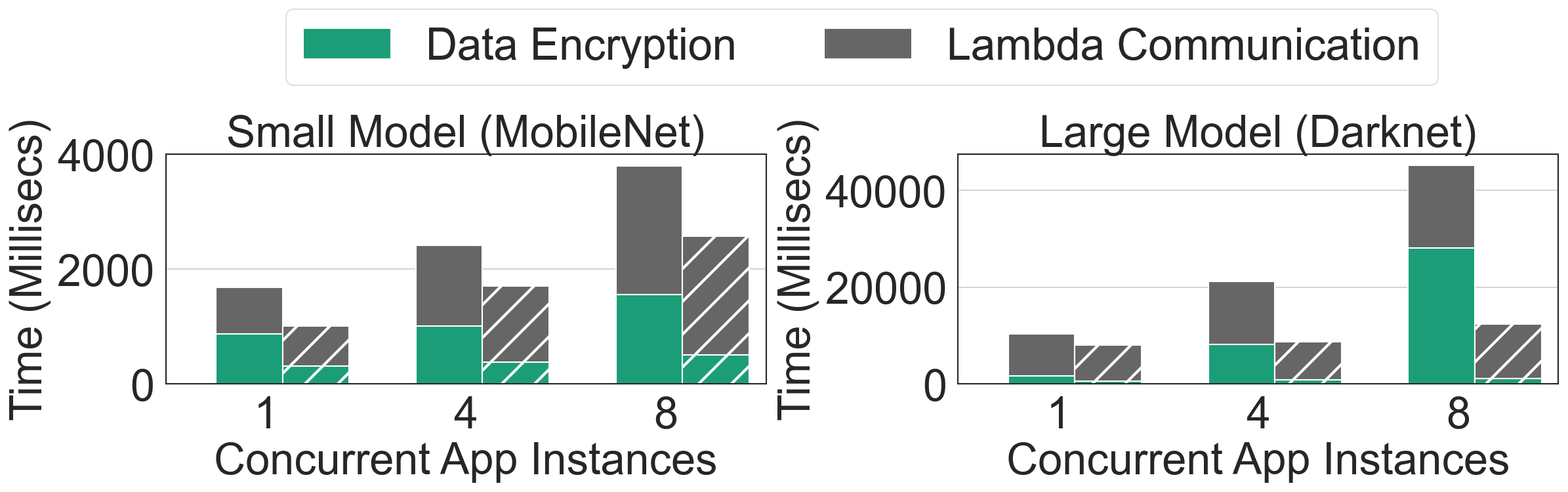}
    \caption{End-to-end latency for \SSIoT hubs to execute apps on the remote AWS Lambda.
    We compare different hardware used for \SSIoT hubs -- Jetson Nano (hatched bars) and Raspberry Pi (plain bars).
    Both platforms experience slowdowns as the number of apps increases.
    }
    \label{fig:eval-hub-benchmark}
\end{figure}
\paragraph{Hub Encryption and Communication Overhead.}
In addition to running the application on FaaS platforms, \SSIoT also requires the local hub to fully encrypt users' private data in order to prevent unauthorized access by malicious apps and third parties. 
User data is encrypted by session keys, and these keys are further encrypted with a specific app key so that only the authorized apps can access each session data (\Cref{sec:encryption-design}).
This encryption overhead is applicable to all apps.
Even if the opportunistic offloading strategy decides to run the app locally, encryption is still necessary for the KMS-based access control.

\Cref{fig:eval-hub-benchmark} shows the relative overhead comparison between data encryption and app execution on Lambda. 
We implement \SSIoT hub on both Raspberry Pi and Jetson Nano.
The major difference is that Raspberry Pi lacks encryption hardware support (e.g., AES)~\cite{rpi-no-crypto-extension} in its ARM processor.
In both cases, the \SSIoT hub spends a considerable amount of time encrypting private data before transmitting it to the target app, often comparable to the Lambda communication time (including time spent on remote app execution). 
Another finding is that, as the number of concurrent applications increases, both encryption and cloud communication tasks experience slowdowns. This is because they start to compete for limited local hub resources (4 cores, 1-4 GB RAM).
Further, data decryption overhead depends on the KMS deployment location. We compare KMS deployments on an AWS EC2 instance in the user's private cloud as well as on the local \SSIoT hub. Our experiments show that key decryption latency in local KMS on \SSIoT hub is ~4.7x slower than cloud deployment (206ms vs. 976ms).

\subsection{Scalability and Concurrent Executions}
\label{sec:eval-scalability}

\begin{figure*}[t]
    \centering
    \small
    \subfloat[Image Classification  - MobileNet.]{
        \includegraphics[width=0.55\columnwidth]{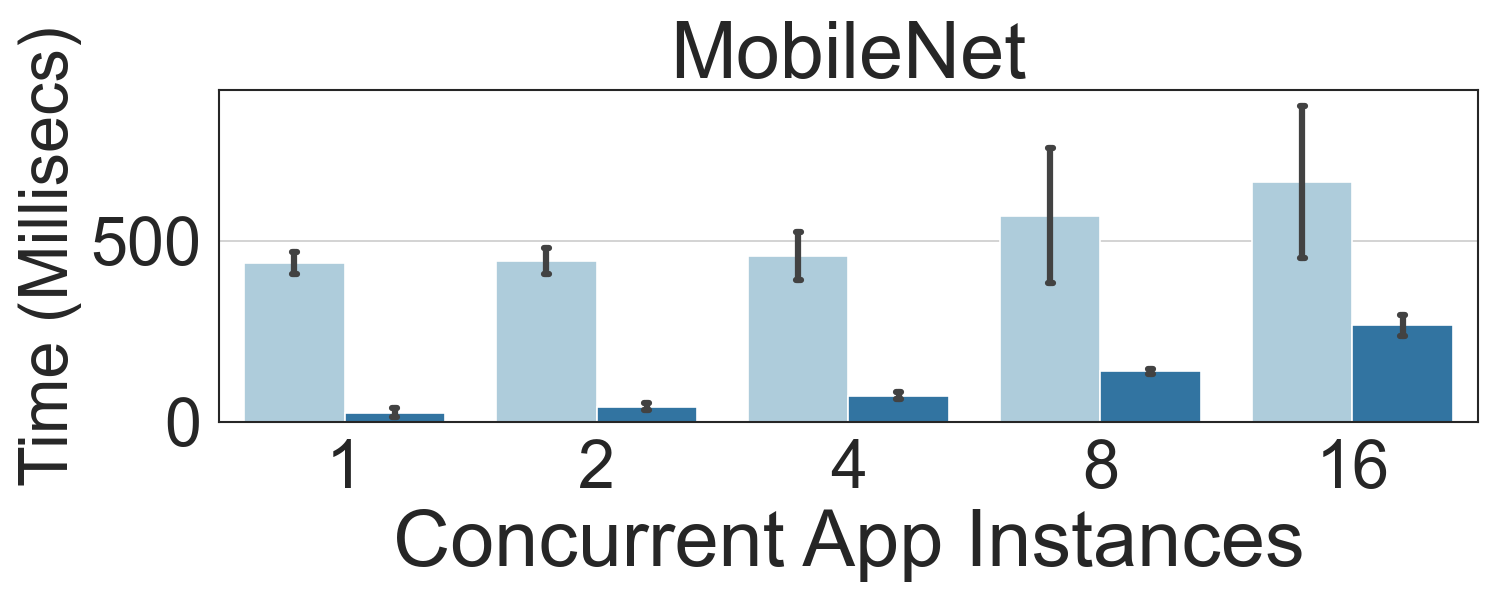}
        \label{fig:eval-scale-mobilenet}
    }
        \subfloat[Image Classification  - DenseNet.]{
        \includegraphics[width=0.55\columnwidth]{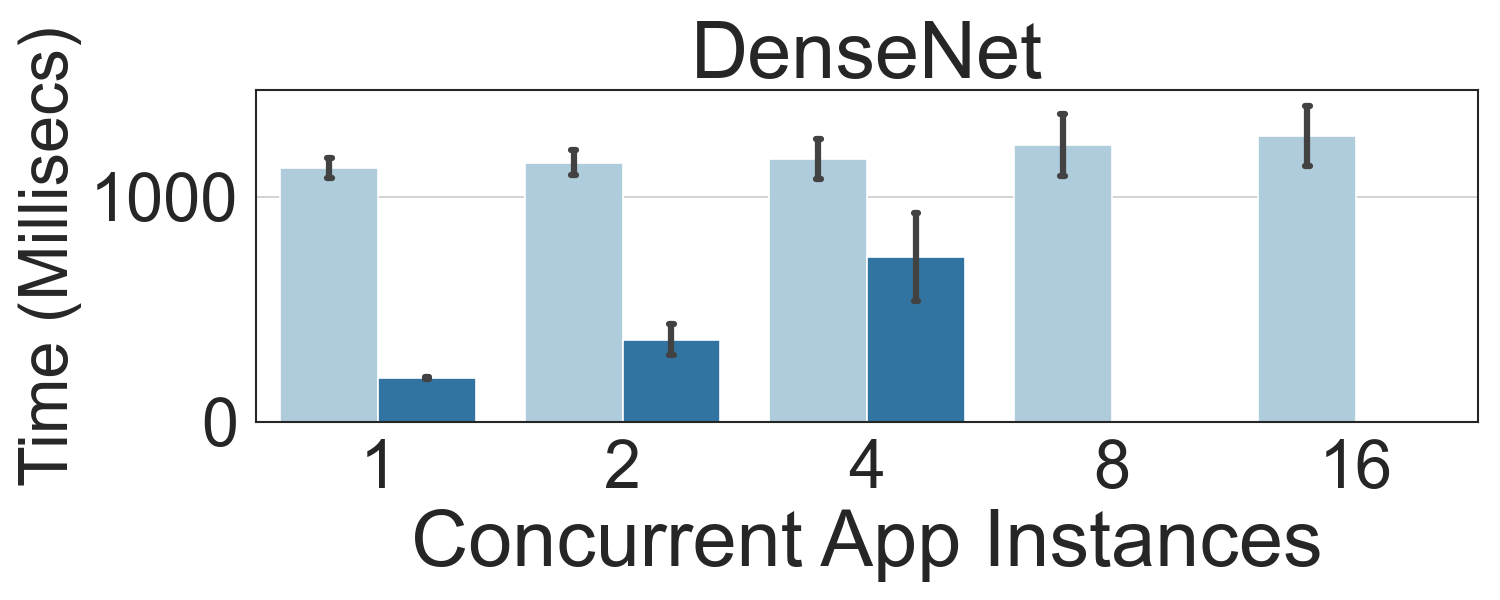}
        \label{fig:eval-scale-densenet}
    }
        \subfloat[Object Detection  - Darknet.]{
        \includegraphics[width=0.55\columnwidth]{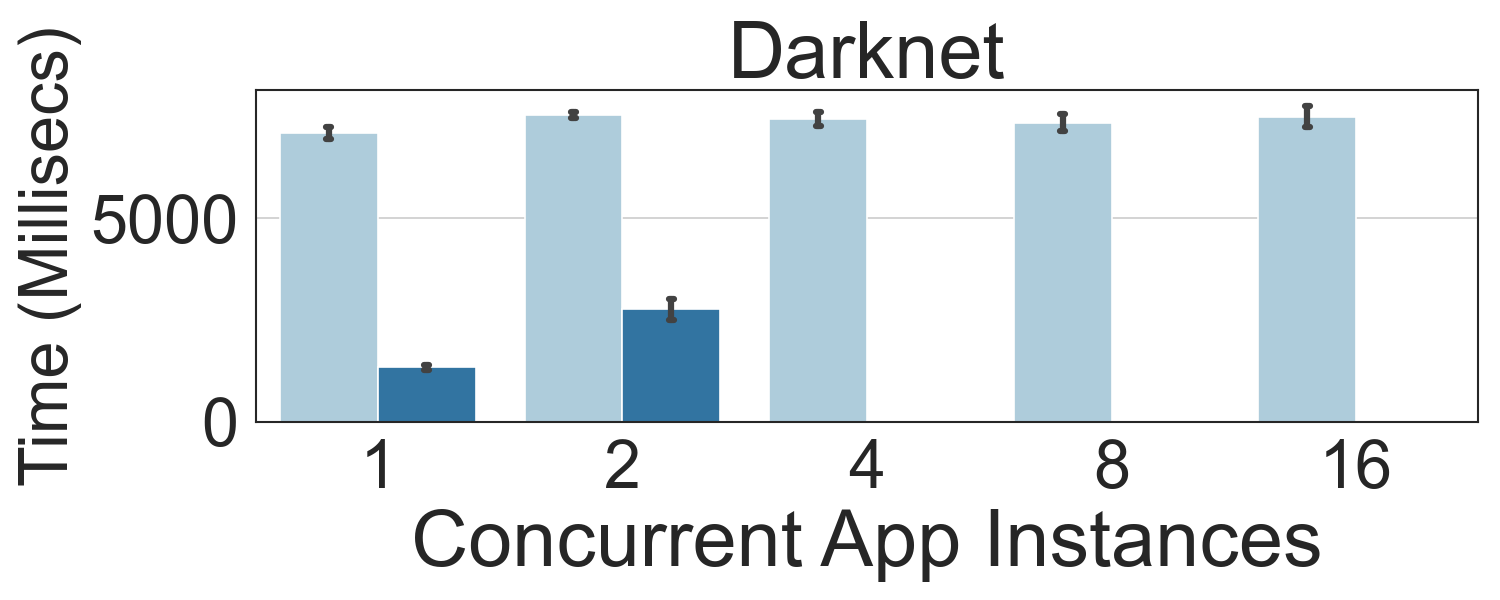}
        \label{fig:eval-scale-darknet}
    }
    \subfloat{
        \raisebox{4mm}{
            \includegraphics[width=0.22\columnwidth]{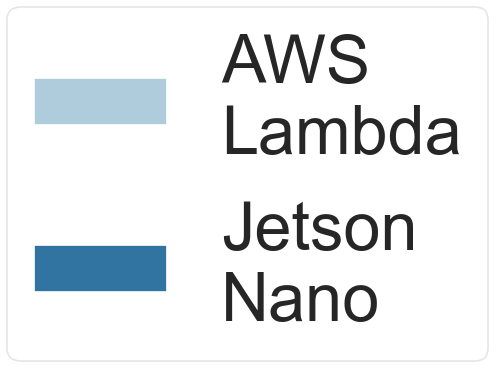}
        }
    }
    \caption{Average latency with standard deviations of concurrent application executions.
    Even with hardware accelerators, Jetson struggles to execute multiple models in parallel (e.g., 2x for large Darknet and 4x for medium DenseNet models) before crashing with depleted resources. 
    Meanwhile, AWS Lambda offloading observes a minor latency increase due to additional communication overhead on the hub.
    }
    \label{fig:scalability-examples}
    \vspace{-4mm}
\end{figure*}

One major advantage of \SSIoT over local devices is the scalability to handle bursty workloads. 
Running one single embedded device like Jetson Nano locally may seem to be an affordable solution for simple home deployment.
However, the data produced by smart devices is often bursty since events from different devices can be correlated due to them being triggered by some human activity~\cite{oconnor2019homesnitch}. 
If multiple events happen all at once, a single local device cannot process them promptly.
\Cref{fig:scalability-examples} shows the capability of a Jetson for various deep learning models. 
Even though the Jetson has an embedded AI accelerator, it cannot scale to more than 4 (e.g., for Image Classification -- DenseNet) or 2 (e.g., for Object Detection -- Darknet) concurrent instances for many models due to memory limits. Ultimately, this limitation in the computational capabilities of local-only \SSIoT hubs will limit what home automation applications' and the complex ML models supporting them can be used.

In contrast, \SSIoT's vision of offloading to FaaS instances alleviates this bottleneck. The \SSIoT hub can initiate multiple inference requests to the FaaS platform. The platform will launch concurrent application instances to process them in parallel. We no longer face the memory and processing limitation to run complex deep learning models at the same time.
Instead, another practical limitation arises on the \SSIoT hub's encryption side. As shown in \Cref{fig:eval-hub-benchmark}, executing multiple data encryption processes on a hub can itself lead to an increase in encryption latency. However, this limitation is much more modest, and it does not restrict our capability of running a large number of applications that are offloaded to the cloud FaaS, as compared to a local-only approach.

\begin{figure}[t]
    \centering
    \small
        \includegraphics[width=0.9\columnwidth]{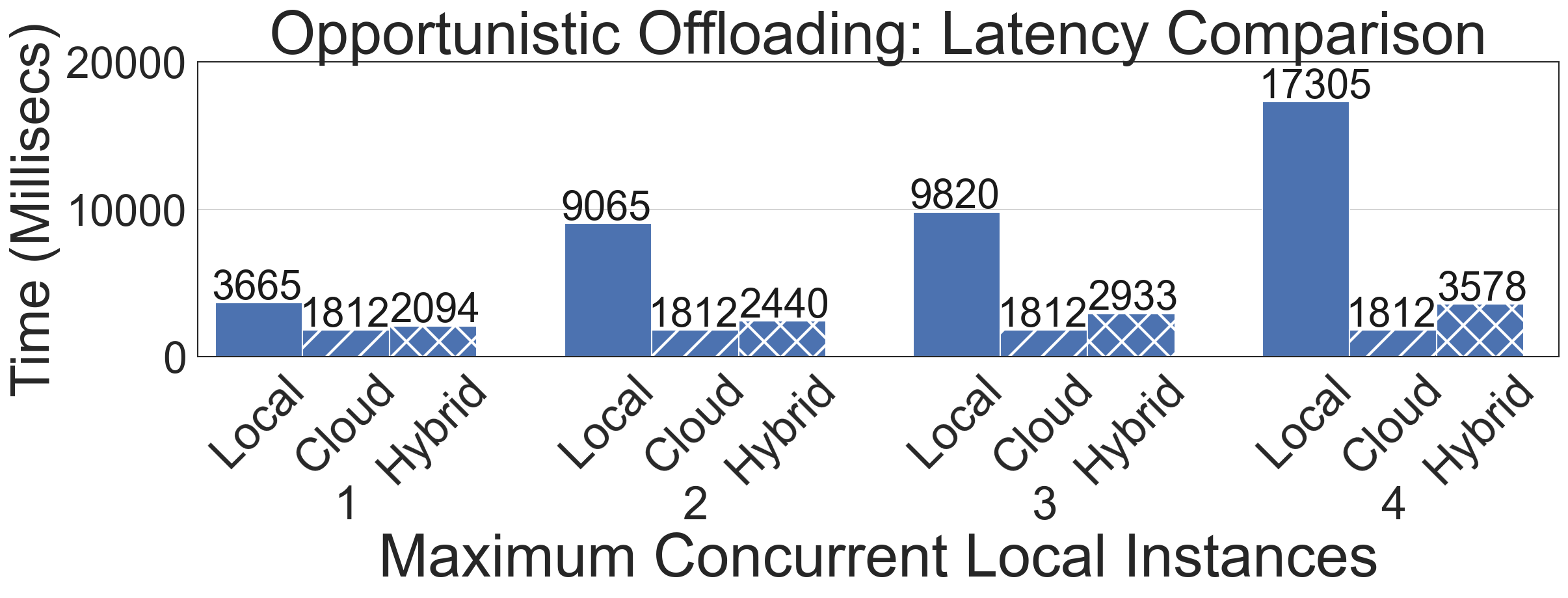}
    \caption{Average latency of \SSIoT's opportunistic offloading and static assignments. 
    Opportunistic offloading (``hybrid'') improves latency over the local-only approach and saves cost over offloading everything to the cloud.
    }
    \label{fig:eval-offload}
    \vspace{-2mm}
\end{figure}

\paragraph{Opportunistic Offloading.}
To measure the effect of opportunistic offloading, we benchmark it to compare with running everything locally and purely Lambda offloading.
We limit the number of maximum local app instances ranging from 1 to 4 in the local-only experiments.
For opportunistic offloading, we launch the same number of local instances along with a cloud offloading handler. It assigns incoming requests to local instances if they are available and offloads them otherwise.
\Cref{fig:eval-offload} compares results of average latency.
As the number of local instances increases, each request experiences slowdown.
The local-only approach sees the most significant impacts (4.7x latency increase as the number of concurrent local instances grows from 1 to 4). 
In comparison, the opportunistic offloading strategy, labeled as ``hybrid'' in the figure, significantly outperforms the local-only approach (up to 79\% latency reduction when the max local instances are set as 4).
More importantly, this strategy processes 15\%-20\% requests locally, saving a portion of cloud invocation costs while maximizing the utilization of free resources.

\subsection{Cost Comparison}
\label{subsubsection:cost_analysis}
\label{sec:eval-cost-analysis}

\begin{table}[t]
    \centering
    \small
	\subfloat[][Baseline.]{
	    \begin{tabular}{c|c|c|c}
        \hline
        Deployment Type & Power & Upfront Cost &  \$ / Month  \\
        \hline
        \hline
        Raspberry Pi & 10W & \$30  & 0.86 \\
        Jetson Nano & 10W & \$100 & 0.86 \\
        Local Desktop & 100W &  $>$\$200 & 8.64 \\
        AWS Reserved & - & - &  5.04  \\
        AWS On-demand  & - & - &  8.53 \\
        \hline
        \end{tabular}
        \label{tab:cost-baseline}
	}
\\
\subfloat[][SSIoT Cost.]{
	    \begin{tabular}{cc|S[table-format=5]|S[table-format=5]}
        \hline
        \multirow{2}{*}{App Type} & \multirow{2}{*}{Model}  & \multicolumn{2}{c}{Requests / \$} \\
    \cline{3-4}
    & & {Cold} & {Warm} \\
        \hline
        \hline
     \multirow{2}{12mm}{Image Classification}  & \texttt{MobileNet}       & 14245 & 65789 \\
      &  \texttt{DenseNet}      &  7133 & 22124 \\
        \hline
      \multirow{2}{12mm}{Object Detection} & \texttt{Darknet} & 1851  &  2896 \\
    &    \texttt{SSDMobilenet}   & 4163  &  7133 \\
        \hline
        \end{tabular}
        \label{tab:cost-ssiot}
	}
	\caption{Cost comparison of \SSIoT and alternatives. We use hourly rates of \texttt{t2.micro} AWS EC2 instance and electricity rates of \$0.12/kWh for baseline. Since cold starts in \SSIoT take significantly longer, they incur higher invocation costs.}
	\label{tab:cost_analysis}
\end{table}

A key goal of \SSIoT is to provide secure remote computation on FaaS platforms while minimizing costs for practicality. 
Multiple factors may affect the cost of using cloud resources, such as the cloud vendor used and where the machines are located. We report costs based on AWS Lambda's pricing while writing this paper, choosing a region closest to us.

\paragraph{AWS Lambda Cost Model.} 
AWS Lambda service charges users based on (i) the number of total invocations to AWS lambda (Requests Costs) and (ii) resource usage per each invocation (Duration Costs).
For `Requests Costs', one million invocations cost \$0.20, which is  \$$2.0 \times 10^{-7}$ per invocation.
Then, the `Duration cost' is a function of the execution time (in seconds) and the memory required (in GB). The rate is \$$1.66667 \times 10^{-5}$ per GB-seconds. We set the maximum configurable memory capacity (3 GB), which gives us \$$5.0 \times 10^{-5}$ per second as the duration cost component. 

\paragraph{\SSIoT is Cost Effective.} 
\Cref{tab:cost_analysis} compares \SSIoT with a local-only alternative.
As a baseline, \Cref{tab:cost-baseline} lists the upfront and operational costs of running single board computers, hardware accelerators, dedicated desktops, and reserved cloud VMs. Running dedicated VMs can cost between \$5.04 to \$8.65 per month, while purchasing multiple low-power embedded devices to support concurrent execution of bursty workloads requires significant upfront costs. The operational costs of local devices are calculated by the average US electricity rates and their power consumption.
In contrast, \Cref{tab:cost-ssiot} demonstrates the capability of \SSIoT.
With a single hub to offload computations, spending \$1 per month can provide as many as 98,039 image classification and 7,133 object detection inferences. 
To put these numbers into context, a smart application like home surveillance monitors can upload one picture every 30 seconds to analyze image contents continuously at the cost of \$1/month (using image classification models) or \$12/month (with object detection to retrieve bounding boxes of all objects), competitive pricing considering existing services charge users as much as \$25/month to process surveillance footage in vendor's cloud~\cite{simplisafe-pricing}.
Of course, many applications do not need continuous execution. As we will show in \Cref{sec:eval-case-study}, more realistic app examples can cost significantly less.

\subsection{Case Study: Integrating Smart Doorbell}
\label{sec:eval-case-study}

To evaluate \SSIoT in a real world scenario, we implemented an end-to-end smart doorbell device (similar to the closed-source Amazon Ring \cite{ring}).
We integrate this device as a \emph{Thing} into the \SSIoT-modified OpenHAB.
With our integration solution, existing OpenHAB-compatible devices can also leverage \SSIoT's new bindings for functionalities such as object detection and image classification. Therefore, there could potentially be many smart applications to be implemented with OpenHAB's IFTTT-rule style programming interface and \SSIoT bindings. However, our smart doorbell application can demonstrate \SSIoT's benefits in cost-effectiveness and realize our vision for building privacy-preserving IoT services  resembling real-world use cases.

To show the simplicity of customizing user applications with IFTTT-style trigger-action rules, we implement this example in \Cref{fig:rule_example}: if a camera detects a person, it sends a text notification to the user. 
Users can pick from a variety of detection models to balance between costs and accuracy.
We conservatively estimated the costs for different models based on a high frequency of app interactions.
We assume 50 invocations per day, lasting 10 minutes each. This sums up to 8.3 hours of continuous monitoring and inference for a smart doorbell every day.
To reduce the invocation frequency and costs, users can deploy a motion sensor to reduce false triggers, or they can restrict the active time windows.
Even without these optimizations, running this smart doorbell application with all-cloud invocations will cost less than \$10 a year from the cloud service providers' billing.

\section{Related Work}
\label{subsection:related_work}

\paragraph{IoT Computation Offloading.}
Private IoT applications can be enabled using a centralized in-house IoT hub like HomeOS \cite{homeos:dixon:2012}, OpenHab \cite{openhab}, and HomeAssistant \cite{home-assistant}. These systems operate locally without requiring internet access, but they are limited in the IoT applications they can support. In contrast, commercial IoT devices such as Nest Camera \cite{nest} directly connect to the cloud via WiFi but are typically unable to function without Internet access \cite{Spying:Apthorpe:2017}.

Recently, edge-cloud hybrid systems have emerged to overcome the limitations of local-only and cloud-only IoT systems. Offloading solutions such as MAUI \cite{MAUI} and NeuroSurgeon \cite{Neurosurgeon} mainly focus on partitioning monolithic mobile applications to the cloud to optimize execution latency and energy usage on mobile devices. SmartThings \cite{SmartThings} implements a monolithic IoT architecture that integrates its own IoT hub and cloud services to support various IoT applications. AWS GreenGrass \cite{aws-greengrass} or Azure IoT \cite{azure-iot} and other research systems such as Beam \cite{Beam:Shen:2016,Beam:Singh:2015} and Steel \cite{Noghabi:Steel:2018} aim to strategically leverage local and remote resources. 
However, these cloud offloading solutions still require users to trust a centralized third-party (e.g., the cloud apps), leading to similar challenges as the current manufacture's app services. 
Instead, a key goal of \SSIoT is to eliminate the need for trusted third parties and enable users to fully control their data and directly manage the local and cloud components of IoT apps themselves. This privacy goal is orthogonal to previous efforts in offloading improvements, though we can leverage their insights to provide better services within \SSIoT.

\paragraph{IoT Private Data Protection.}
There are efforts to protect private data inside various platforms by deploying applications inside trusted secure enclaves \cite{elgamal2020serdab, hunt2018chiron}. However, these approaches require careful instrumentation to pre-existing frameworks such as Se-Lambda \cite{SeLambda}. Also, running applications inside secure enclaves impose significant performance overhead and practical constraints on data sizes and application memory usages \cite{Slalom}.

Several works have proposed approaches to prevent private data from leaving the user's local environment.
Jayaraman et al.~\cite{jayaraman2017privacy} propose special privacy-preserving functions for IoT device data collection.
Karl~\cite{yuan2022karl} and Peekaboo~\cite{jin2022peekaboo} propose new privacy policy enforcement mechanisms in a local-only IoT execution environment.
Davies et al. \cite{iot_chasm:2016} and Wang et al. \cite{Wang:RTFace:2017} proposed interesting system architectures to have a mediator between edge devices and remote cloud-based IoT hubs. Those mediators are virtual machines hosted inside cloudlets \cite{Satyanarayanan:Cloudlet:2017} which are local, smaller-scale datacenters located in the vicinity of the edge devices. The mediators get the privacy-sensitive, raw data from sensors and denature (or anonymize) the data before sending the data to the cloud service. Even though these approaches can prevent remote cloud operators from getting raw data, the trust is shifted to the cloudlet operators instead. Cloudlets systems are not readily available yet, but in the future, the core components of SSIoT can potentially be run on cloudlets instead of FaaS platforms.

\paragraph{Serverless Offloading.}
Several research efforts have also explored how to leverage serverless computing to augment local devices' limitations~\cite{cicconetti2019low,DIY,cicconetti2020architecture, executionmodelforserverlessfunctionsatedge}.
Our design choice of using the FaaS paradigm was partly inspired by 
DIY~\cite{DIY}, which explores executing private web services such as E-mail on serverless platforms. 
\SSIoT extends these ideas into IoT domains and addresses new challenges in practicality and machine learning workloads commonly present in IoT apps.

\section{Discussion and Limitations}

\paragraph{Alternative Approaches for IoT Privacy.}
We acknowledge that end user privacy for IoT devices can also be enforced in other ways. The most common one is for users to assess privacy risks by reading the vendor's privacy policies and deciding whether or not to use the device. However, prior work has shown that users don't read these policies, and are incapable of assessing the actual privacy risks hidden behind pages of legalese. Ultimately the user still does not know whether to believe the vendor's intention of what they do with their data. Another approach is to introduce regulations (e.g., General Data Protection Regulation \cite{GDPR}). These can be effective in providing guidelines and mechanisms to audit the privacy practices of vendors. Still, they are not a panacea since vendors can still look to evade them. \SSIoT, in contrast, is aimed at providing users an alternative approach where they are completely in control of their IoT device data and the computations performed on it, even if offloaded to the cloud.

\paragraph{Restricting Data Sharing across Applications.} We focus on offloading complex machine learning applications (DNNs) with \DIYL. However, there are also other types of privacy-invasive IoT applications that require aggregating data among different users. Such applications may no longer work due to the strict data isolation provided by \SSIoT, where different apps from multiple homes do not share data with one another.

\section{Conclusion}

In this paper, we propose \SSIoT, a novel offloading framework that addresses privacy threats around the prevailing model of IoT services by allowing users to deploy and manage their own IoT applications. It is composed of a local hub, a remote runtime that runs on a public FaaS environment, a toolchain to automate application packaging and deployment, and a key management service to assist with secure data exchange between the local hub and remote application runtime. It utilizes an opportunistic offloading strategy that brings benefits of both local-only and cloud-based solutions to users with full control over their data. Our evaluation with a comprehensive suite of machine learning models and a real-world use case of a smart doorbell shows that \SSIoT is cost-efficient, dynamically scalable, and is a practical solution to address privacy challenges in future smart homes.

\bibliographystyle{plain}
\bibliography{bibliography}

\appendix
\section*{Appendix}

\begin{figure*}[!ht]
    \centering
    \includegraphics[width=1.98\columnwidth]{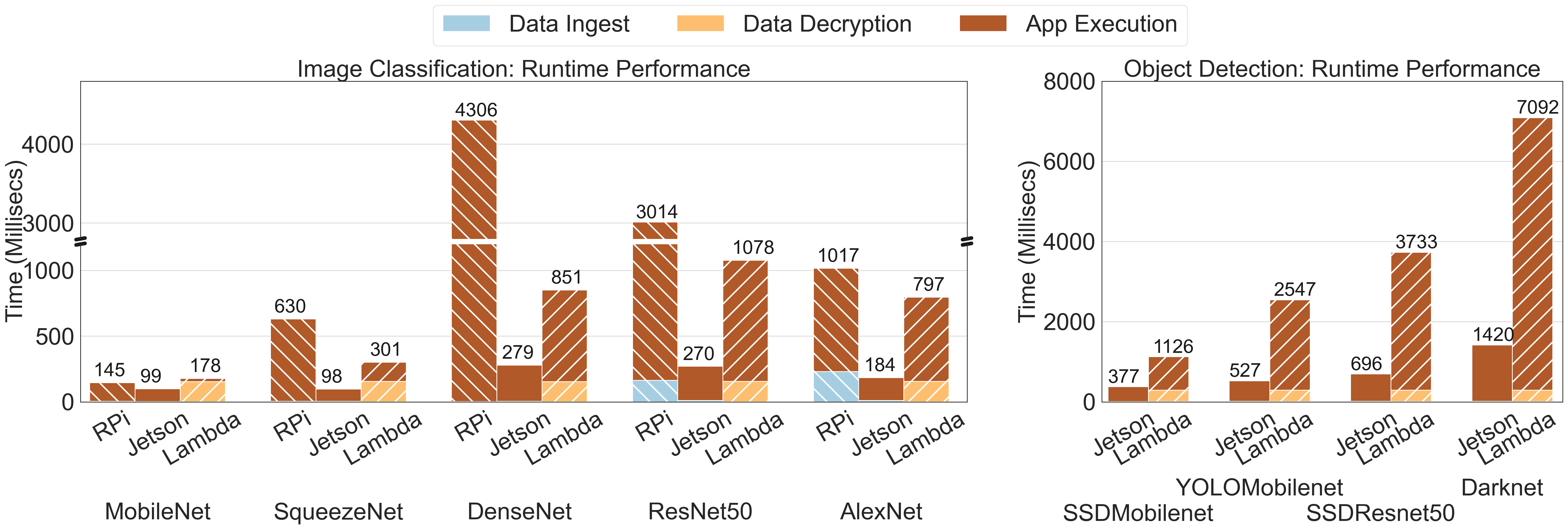}
    \caption{Average inference latency across different platforms and models. 
    Raspberry Pis cannot perform object detection due to limited resources.
    Moving apps to lambda reduce latency for single board computers (RPi), but the additional network communication overhead overshadows the processing benefit when compared to hardware AI accelerators (Jetson Nano).
    }
    \label{fig:appendix-eval-function-latency}
\end{figure*}

\begin{figure*}[!ht]
    \centering
    \includegraphics[width=1.98\columnwidth]{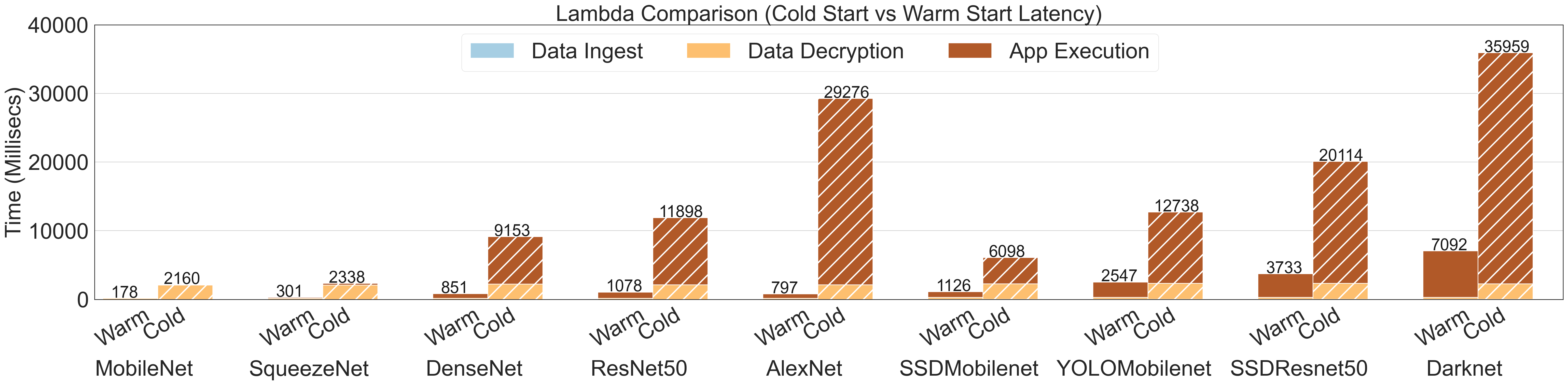}
    \caption{Average inference latency between cold and warm AWS lambda instances.
    Cold states drastically slow down app execution (up to $>$36x) due to delays in app instantiation.
    }
    \label{fig:appendix-eval-cold-start-latency}
\end{figure*}

\section{Complete \SSIoT Workflow}
\label{sec:appendix-workflow}

In this section, we outline how a \SSIoT hub deploys an IoT app (\S \ref{subsubsection:app_deploy}) and invokes the remote part of the app (\S \ref{subsubsection:app_invoke}). We show the step-by-step procedures in Figure \ref{fig:procedure}. 

\subsection{\SSIoT Toolchain: Deploying App}
\label{subsubsection:app_deploy}

We now describe the operation details of our \emph{\SSIoT toolchain}. The toolchain is responsible for deploying an application that users download from a repository of \SSIoT bindings.

\emph{Initial Setup (D1):} Before deploying any application to the cloud, \SSIoT toolchain first obtains the list of available key management services (\SSIoT KMS) and fetches their public keys ($K_{M}$). \SSIoT toolchain can decide which KMS to use for app deployment at this stage.

\emph{Generating \& Registering App-specific Key Pair (D2-3):} Once a local hub or a user specifies an application to offload, \SSIoT toolchain first fetches the application from the internet or local storage if the application code is already downloaded. Then, the toolchain generates a per-application RSA key pair ($K_{A}$, $K_{A}^{-1}$). The key pair is mainly used by the \SSIoT hub to sign and encrypt private data sent to the serverless platform and the \SSIoT KMS. The \SSIoT toolchain sends the application's private key ($K_{A}^{-1}$) to the KMS.
This allows the KMS to identify and decrypt the private data key from \SSIoT hub. 

\emph{Preparing \& Deploying App Package (D4-5):} \SSIoT toolchain finally build a package that includes \texttt{app\_binary}, identity of KMS ($M$) and the app's private key ($K_{A}^{-1}$). When building a package, the app function is packaged and compiled by the toolchain to ensure \SSIoT libraries are properly merged with the downloaded application.

\subsection{\SSIoT Runtime: Invoking App}
\label{subsubsection:app_invoke}

Once the application is successfully deployed to the serverless platform, the local \SSIoT runtime can invoke the deployed application.

\emph{Preparing Invocation (I1-2): } A \SSIoT local hub collects data (e.g. image or audio) from the local environment and generates a random symmetric \emph{data key} $K$ in \texttt{I1}. In \texttt{I2}, the private data is encrypted with the app's public key ($K_{A}$) and then encrypted with the symmetric key $K$. In addition, the symmetric key itself is also encrypted asymmetrically by the app's public key ($K_{A}$) and the KMS's public key ($K_{M}$). Note that the data key is not pre-installed in any of the entities in our system and can be generated whenever the local \SSIoT hub wants to do so (e.g., due to suspicious KMS behavior, etc.).

\emph{Invoking Application inside Serverless Platform (I3-4):}
Given plaintext \texttt{data}, \SSIoT runtime encrypts the data with app-specific public key $K_{A}$ and then encrypts it with the data key $K$.
To share $K$ with the remote serverless platform runtime, the local runtime also sends out $K$ signed and encrypted with the app's public key $K_{A}$ and KMS's public key $K_{M}$ (\texttt{encrypted\_key}).
Note that $data$ and $K$ are not exposed to serverless platform' operators even when the payload is traveling inside the serverless platform's internal infrastructure (\texttt{I4}).

\emph{Acquiring Data Key (I5-7):} By the time a function runtime ($R$) receives \texttt{encrypted\_data}, it cannot decrypt the payload because it does not have $K$. To get the data key $K$, it sends out \texttt{encrypted\_key} to the \SSIoT KMS and gets $K$ in plaintext over a secure TLS channel. The KMS additionally verifies the identity of data origin when decrypting \texttt{encrypted\_key}.

\emph{Running Application (I8):} The remote \SSIoT runtime now decrypts \texttt{encrypted\_data} and run the application inside an isolated sandbox (\texttt{I7}). In this sandbox, the application is isolated from the file system and network. We enforce file accesses control inside the instance so that an application cannot rewrite the \SSIoT runtime logic. Similarly, we also restrict network access to prevent the application from sending out private data outside of \SSIoT's control other than the communication between KMS and the runtime.

\emph{Returning Results (I9-10)} Finally, the processed response of the application is encrypted with data key $K$ to protect it during transmission back to the local hub ($H$).

\section{Additional Evaluation Results}
\label{sec:additional-eval-results}

\Cref{fig:appendix-eval-function-latency} presents average inference latency for executing a variety of machine learning models in \SSIoT. We compare \ssiot offloading with local-only approaches on different hardware platforms (Raspberry Pi and Jetson Nano).
\Cref{fig:appendix-eval-cold-start-latency} compares the average latency between invoking applications on a cold vs warm  remote FaaS instance. 
\end{document}